\documentclass[conference]{IEEEtran}
\IEEEoverridecommandlockouts
\usepackage{cite}
\usepackage{amsmath,amssymb,amsfonts}
\usepackage{algorithmic}
\usepackage{graphicx}
\usepackage{textcomp}
\usepackage{fancyhdr}
\usepackage{enumitem}
\usepackage{comment}
\usepackage{multirow}
\usepackage{cases}
\usepackage{multicol}
\newtheorem{theorem}{Theorem}[section]
\newtheorem{corollary}[theorem]{Corollary}
\newenvironment{proof}[1][Proof.]{\begin{trivlist}\item[\hskip \labelsep {\bfseries #1}]}{{\hfill$\blacksquare$\bigbreak} \end{trivlist}}
\newenvironment{remark}[1][Remark]{\begin{trivlist}\item[\hskip \labelsep {\bfseries #1}]}{\end{trivlist}}

\def\BibTeX{{\rm B\kern-.05em{\sc i\kern-.025em b}\kern-.08em
    T\kern-.1667em\lower.7ex\hbox{E}\kern-.125emX}}
\makeatletter
\newcommand*\titleheader[1]{\gdef\@titleheader{#1}}
\AtBeginDocument{%
  \let\st@red@title\@title
  \def\@title{%
    \bgroup\itshape\small\raggedright \@titleheader\vskip-\headrulewidth\hrule\par\egroup
    \vskip1.5em\st@red@title}
}
\makeatother

\title{On the Complexity of the K-way Vertex Cut Problem\\
}

\titleheader{International Conference on Artificial Intelligence and Information Technology, ICA2IT'19}

\begin{document}

\author{\IEEEauthorblockN{Mohammed LALOU}
\IEEEauthorblockA{\textit{Department of M.CS, Institute of Science \& Technology} \\
\textit{University Center A. Boussouf of Mila}\\
Mila, Algeria \\
mohammed.lalou@centre-univ-mila.dz}

}

\maketitle

\begin{abstract}
The K-way vertex cut problem consists in, given a graph $G$, finding a subset of vertices of a
given size, whose removal partitions $G$ into the
maximum number of connected components. This problem has many applications in several areas. It has been
proven to be NP-complete on general graphs, as well as on split and planar
graphs. In this paper, we enrich its complexity study with two new results. First, we prove that it remains NP-complete even when restricted on the class of
bipartite graphs. This is unlike what it is expected, given that the K-way vertex cut problem is a generalization of the Maximum Independent set problem which is polynomially solvable on bipartite graphs. We also provide its equivalence to the well-known problem, namely the \textit{Critical Node Problem} (\textit{CNP}), On split graphs. Therefore, any solving algorithm for the \textit{CNP} on split graphs is a solving algorithm for the \textit{K-way vertex cut problem} and vice versa. 
\end{abstract}

\begin{IEEEkeywords}
Vertex separator, critical nodes, graph connectivity,
bipartite graphs, split graphs, NP-completeness.
\end{IEEEkeywords}

\section{Introduction}

Given an undirected graph $G=(V,E)$, where $V$ is the set of
vertices and $E$ the set of edges, and an integer $k$, we ask for
a subset $S \subseteq V$ of $k$ vertices whose deletion maximizes
the number of connected components in the induced subgraph $G[V
\setminus S]$. Note that $G[V
\setminus S]$ denotes the subgraph induced by $V \setminus S$. This problem is known as the \emph{K-way vertex cut problem} \cite{ber14,shen12}. Its recognition version can be stated as
follows. Let $c(G)$ denotes the number of connected
components in $G$.\vspace{0.3cm}

\noindent\textbf{K-way Vertex Cut Problem (KVCP)}
\vspace{0.2cm}

\noindent\textbf{Instance:} A graph $G = (V, E)$, and an integer
$k$.\vspace{0.05cm}

\noindent\textbf{Question:} Is there a subset of
vertices $S \subseteq V$, where $|S| \leq k$, the deletion of which satisfies $c(G,[V\setminus S])
\geq K ?$ where $K$ is an integer.\vspace{0.3cm}

The objective is to find a subset $S \subseteq V$ of at most $k$
vertices, the deletion of which partitions the graph into at least $K$
connected components. This problem is the vertex-version of the
well-known \emph{Minimum k-cut problem}
\cite{gol94,kar96,kam06,sar95,dow03}, where we ask for deleting a
set of edges instead of vertices, with the purpose of maximizing
the number of connected components in the induced graph. Note that
the number of connected components in a graph can be computed in
linear time using either breadth-first search or depth-first
search algorithm \cite{cor09}.

The \emph{K-way vertex cut problem} has been proven to be
NP-complete on general graphs \cite{shen12} through a reduction
from the \emph{Maximum Independent Set problem} (\emph{MIS}). Indeed, we
can easily see that any subset of vertices whose deletion
separates the graph into at least $K$ components identifies an
independent set of size at least $K$. Accordingly, the \emph{K-way
	vertex cut problem} on $G$ is a natural generalization of the
\emph{MIS} on $G$. Conversely, the \emph{MIS} is the particular case
of the \emph{K-way vertex cut problem} where the connected
component size has to be equal to one. The two problems are
equivalent if $k\geq |V\setminus I|$ where $I$ is the maximum
independent set of $G$.

One can hope that the \emph{K-way vertex cut problem} becomes
polynomial on classes of graphs for which \emph{MIS} is
polynomially solvable. However, this is not the case for the class
of bipartite graphs. In this paper, we prove that the \emph{K-way
	vertex cut problem} remains NP-complete even on this class of
graphs. While for the class of split graphs, we
provide an equivalence between the \textit{KVCP} and \textit{CNP}. This allows the \textit{KVCP} to be solved using any algorithms for solving the \textit{CNP}. 

\begin{figure}[htbp]
	\centerline{\includegraphics[scale=0.22]{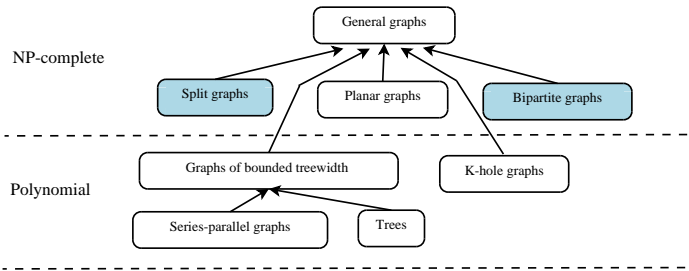}}
	\caption{The complexity of the K-way vertex cut problem on different classes of graphs. Contributions of this paper concern the colored classes.
	} \label{graph_classes}
\end{figure}

Figure \ref{graph_classes} reviews the \textit{KVCP} complexity on different classes of graphs considered in the literature and highlights the contributions of this paper. 


The rest of the paper is organized as follows. We complete this
section by a state-of-the-art of the \emph{K-way vertex cut
	problem}, where we review different works handled this problem in
the literature. Also, we give some definitions we need in the rest
of the paper. In section \ref{sec-bG}, we provide the
NP-completeness proof of the problem on bipartite graphs. In
section \ref{sec-sG}, we deduce its equivalence to the \textit{CNP}, while in section \ref{sec-btw} we deduce its resolvability in polynomial time on weighted graphs of bounded treewidth. We close up the paper by some future
works in section \ref{sec-conc}.

\subsection{Related works}

The \emph{K-way vertex cut problem} can be considered as a
parametrized version of the graph separation problem
\cite{mar06}, where we ask for the vertex-separator set that
partitions the graph into the maximum number of connected
components. As well, it can be considered as a variant of the
\emph{Critical Nodes Detection Problem (CNDP)} \cite{lal17}. This
problem (\emph{CNDP}) consists in finding the subset of vertices
whose removal significantly degrades the graph connectivity
according to some predefined connectivity metrics, such as:
minimizing the pairwise connectivity in the network
\cite{add13,aru09,aru11,sum11}, minimizing (or limiting to a given
bound) the largest component size \cite{she12,shen12,lal15}, etc.
In the case of the \emph{K-way vertex cut problem} the 
metric considered is maximizing the number of connected components.

Although its importance, the \emph{K-way vertex cut problem} has
received  a little attention, in the literature, as expected for such an important problem. On general
graphs, the problem has been shown to be NP-complete
\cite{shen12,ber14}, and NP-hard to be approximated within a
factor of $n^{(1-\epsilon)}$, for any $\epsilon> 0$ \cite{ber14}.
Also, it is W[1]-hard, \emph{i.e.} not fixed-parameter tractable,
with respect to the two parameters, namely the number of deleted
vertices $k$, and the number of connected component in the induced
graph $K$ \cite{mar06}. We recall that when we deal with
parametrized problems, the input instance has an additional part
called \emph{parameters}. The complexity of the problem is then
measured as a function of those parameters, and the problem is
said to be fixed-parameter tractable if it can be solved using
algorithms that are exponential only in the size of these
parameters, while polynomial in the size of the problem instance.

For solving the \emph{K-way vertex cut problem} on general graphs,
a \emph{Mixed-Integer Program} formulation has been presented in
\cite{shen12}, where bounds and validated inequalities for the
proposed formulation have been studied. As well, an evolutionary
framework, that uses two greedy methods embedded within two main
genetic operators, has been presented in
\cite{ari16b}. The two operators, namely reproduction and mutation, are used to repair the obtained
solutions, while the greedy methods are used to guide the search in
the feasible solution space.

Considering the  \emph{K-way vertex cut problem} on particular
classes of graphs, it has been proved to be NP-complete on split
and planar graphs \cite{ber14}. Also it has been shown, by the same authors, that the problem is NP-hard to be approximated on split
graphs \cite{ber14}, while on planar graphs it can be approximated
using a polynomial-time approximation scheme (PTAS) of complexity
$O(nk^2f(\epsilon))$, where $\epsilon> 0$ and $f$ is a function
only depending on $\epsilon$ \cite{ber14}. We note that a PTAS
outputs an approximate solution of value at least $(1- \epsilon)$
times the optimum, and the running time is polynomial in the size
of the problem. Considering the parametrized complexity on these
two classes of graphs, the problem remains W[1]-hard with
respect to parameter $k$ (the number of vertices to be deleted) on split
graphs \cite{ber14}, however on planar graphs, a fixed-parameter
tractable algorithm of complexity $O(nk^{O(k)})$, with respect to
$k$, has been proposed \cite{ber14}.

On trees, $k$-hole and series-parallel graphs, polynomial dynamic
programming algorithms have been developed for solving the problem
with complexity $O(n^3)$, $O(n^{3+k})$ and $O(n^3)$, respectively
\cite{she12}. Also on graphs of bounded treewidth, the problem can
be solved in polynomial-time using a dynamic programming algorithm
with complexity $O(nk^2w^w)$, where $w-1$ is the treewidth
\cite{ber14}.

\emph{Table \ref{maxnum_comp}} summarizes the different results
arisen from studying the \emph{K-way vertex cut problem} on
different classes of graphs.\\

\begin{table*}[!ht]
	\caption{The different results obtained from studying the
		\emph{K-way vertex cut problem} on different classes of
		graphs.}
	\begin{center}
		\begin{tabular}{|c|c|c|c|}
			\hline
			\textbf{Graph class}& \textbf{Complexity} & \textbf{Solving approach} & \textbf{Time}\\
			
			\hline\hline General graphs & \multirow{4}{*}{NP-complete \cite{shen12,ber14}} & Genetic algorithm \cite{ari16b} & NC\\
			\cline{1-1}\cline{3-4} \multirow{2}{*}{Planar graphs} &
			\multirow{4}{*}{} & PTAS \cite{ber14}&
			$O(nk^2f(\epsilon))$\\
			\cline{3-4}
			
			& & FPT \cite{ber14}& $O(nk^{O(k)})$\\
			\cline{1-1}\cline{3-4} Split graphs & &\multicolumn{2}{c|}{$\setminus$} \\
			
			\cline{1-4}
			
			Trees & \multirow{4}{*}{Polynomial \cite{she12}} & \multirow{3}{*}{Dynamic programming \cite{she12}} & $O(n^3)$ \\
			\cline{1-1}\cline{4-4}
			k-hole graphs & &  &  $O(n^{3+k})$ \\
			\cline{1-1}\cline{4-4}
			Series-parallel graphs & &  & $O(n^3)$ \\
			\cline{1-1}\cline{3-4}
			Graphs with bounded $T_w$ & &  Dynamic programming \cite{ber14}& $O(nk^2w^w)$ \\
			
			\hline
		\end{tabular}
		
	\end{center}
	\label{maxnum_comp}
\end{table*}

\subsection{Definitions and notations}

Let $G = (V,E)$ be an undirected graph, where $V$ is the set of
vertices and $E \subseteq V \times V$ is the set of edges. Two
distinct vertices $u$ and $v$ are adjacent (or neighbour) if there
exists an edge $uv \in E$ connecting them. $u$ and $v$ are called
the endpoints of the edge $uv$. The neighbourhood set of a vertex $v
\in V$ is defined as $N(v)=\{u \in V | \{u,v\} \in E\}$. Let
$deg_G(v)$ denote the degree of the vertex $v$, we have
$deg_G(v)=|N(v)|$. 

A \emph{chain} in $G$ is a sequence of distinct
vertices $\{v_1, v_2,\ldots,v_k \}$ such that $v_iv_{i+1}$ is an
edge for each $0 < i <k-1 $.

Given a subset of vertices $S \subseteq V$, $S$ is called an
\emph{independent set} if there are no edges between any pair of
vertices in $S$.  We use $G[S]$ to denote the subgraph of $G$
induced by $S$, and hence $G[V\setminus S]$ denotes the subgraph induced
by $V\setminus S$. Also, we use $c(G,S)$ to denote the number of
connected components in $G[V\setminus S]$ obtained by removing $S$
from $G$. As well, $c(G,A)$ denotes the number of connected components
obtained by deleting a set of edges $A \in E$.

A graph $G=(V,E)$ is a \emph{bipartite graph} if the vertex set
$V$ can be divided into two disjoint subsets $V_1$ and $V_2$, such
that every edge $e \in E$ has one endpoint in $V_1$ and the other
endpoint in $V_2$. Each subset, $V_1$ or
$V_2$, forms an independent set of $G$. $G$ is then denoted $G=(V_1,V_2,E)$, where $n_1
= |V_1|$, $n_2 = |V_2|$ and $n_1 + n_2 = n$. $G$ is said to be a
complete bipartite graph, denoted $K_{n_1,n_2}$, if each vertex in
$V_1$ is adjacent to all vertices in $V_2$. If one of the independent set, $V_1$ or $V_2$, is a clique $G$ is called a split graph.

\section{Bipartite graphs}
\label{sec-bG}

In this section, we consider the \emph{K-way vertex cut problem}
on bipartite graphs. This case is relevant when the network to be
decomposed on connected groups or communities has a bipartite
structure, which is the case, for example, of users vs files in a P2P system, traders vs
stocks in a financial trading system, conferences vs authors in a
scientific publication networks and so on.

In the following, we show that the \emph{K-way vertex cut problem}
remains NP-complete even on this class of graphs. In order to
establish the complexity proof we have first to introduce the
following transformation of the \emph{k-cut problem} on general
graphs \cite{gar79} to the \emph{K-way vertex cut problem} on
bipartite graphs. \vspace{0.3 cm}

\noindent\textbf{\textit{The K-cut problem.}} Given a graph $G=(V,E)$ and
an integer $K$, find a minimal subset of edges $A \subseteq E$,
whose removal partitions the graph into at least $K$ connected
components, \emph{i.e.} such that $c(G, A)\geq K$. This problem is
NP-complete on general graphs \cite{gar79}, and its recognition
version asks whether there exists a cut-edges set $A$ where $|A|\leq B$, for a given bound $B$.

\begin{figure}[htbp]
	\centerline{\includegraphics[scale=0.18]{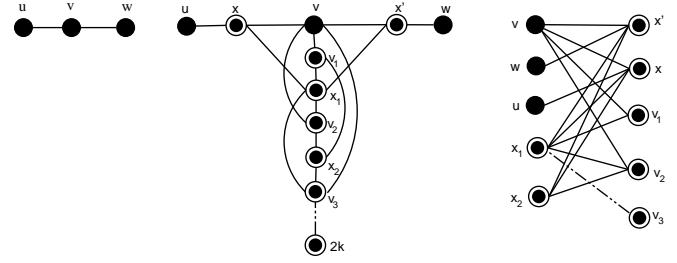}}
	\caption{The reduction \emph{k-cut problem $\propto$ K-way vertex
			cut problem} on bipartite graphs. The added vertices are those
		with circles, and we have $U=\{x,x'\}$.} \label{proof_bipartite}
\end{figure}

We give a polynomial-time reduction from the \emph{k-cut problem}
to the \emph{K-way vertex cut problem}. Given an instance of the
\emph{k-cut problem} on a general graph $G(V,E)$, we define an
instance of the \emph{K-way vertex cut problem} on a bipartite
graph $G'=(V',E')$ as follows:

\begin{enumerate}
	
	\item $G'$ contains all vertices and all edges of $G$, \emph{i.e.}
	$V \subseteq V'$ and $E \subseteq E'$.
	
	\item For each vertex $v \in V$, if $deg_G(v) \geq 2$, we add to
	$G'$ a chain $p_v=\{v_1,\dots,v_k\}$ of $k$ vertices, such that
	$v_1$ coincides with $v$ (see \emph{Figure
		\ref{proof_bipartite}}).

	\item For each edge $uv \in E'$, we add a vertex $x \in V'$ such
	as we replace $uv$ with two new edges $ux, xv \in E'$, \emph{i.e.}
	we replace each edge $uv$ by a chain $\{u,x,v\}$ such that $x$ is
	an added vertex. We denote $U$ the set of all added vertices $x$
	for which $ux,xv \in E'$ and $u,v \in V$.
	
	\item For each chain $p_v=\{v_1,x_1,v_2,x_2,\dots,v_k\}$ we add
	two edges $xx_1, x'x_1 \in E'$ where $x_1\in P_v$ and $xv,vx'$ are
	two edges sharing the vertex $v\in V$ (see \emph{Figure
		\ref{proof_bipartite}}). Also we add edges $vv_i$ where $2\leq
	i\leq k$, and $v_ix_{i+1}$, $x_iv_{i+1}$ where $1\leq i\leq k$ for
	each chain $p_v$.

\end{enumerate}

Note that removing vertices of $V$ from $G'$ does not disconnect
the graph $G'$, and $G'$ becomes disconnected only by removing
vertices of $U$. Also, it is obvious that the transformation can
be done in polynomial time, and the graph $G'$ is bipartite. Now,
we prove the following theorem.\vspace{0,2cm}

\begin{theorem}
	\label{NP-C bipartite} The K-way vertex cut problem is NP-complete
	on bipartite graphs.
\end{theorem}

\begin{proof}
	
	The \emph{K-way vertex cut problem} is in \emph{NP} since given a
	graph, we can compute in polynomial time the number of connected
	components in the induced graph after deleting $k$ vertices. Now
	we prove that the \emph{K-cut problem} on general graphs $\leq_p$
	\emph{K-way vertex cut problem} on bipartite graphs.
	
	Given an instance $I$ of the \textit{K-cut problem} on a general graph
	$G=(V,E)$, we construct an instance $I'$ of the \emph{K-way vertex
		cut problem} on a bipartite graph $G'(V',E')$ as described in
	(1)-(4). We show that $G$ has a cut-edge set $A \in E$ of $k$
	edges such that $c(G,A) \geq K$ if and only if $G'$ has a
	cut-vertex set $S \in V'$ of $k$ vertices such that
	$c(G',V'\setminus S) \geq K$.
	
	First, let $A \subseteq E$ be a solution of $I$, so $A$ contains
	no more than $k$ edges whose deletion disconnects $G$ into at
	least $K$ components. In $I'$, we select the k vertices of S as
	follows: for each edge $uv \in A$, we select from $G'$ the
	corresponding vertex $x \in U$ such that $ux,xv \in E'$.
	
	By deleting the vertices in $S$ from $G'$, no more than $k$
	vertices are deleted $|S|\leq k$ and at least $K$ connected
	components are generated $c(G',V'\setminus S) \geq K$. Hence, $S$
	is a solution of the \emph{K-way vertex cut problem} on $G'$.
	
	Conversely, we prove that if there is a cut-vertex set $S$ of size
	$k$ for $G'$, then we have a cut-edge set of size $k$ for $G$. Let
	$S$ be a solution of $I'$, so $S$ contains a set of $k$ vertices
	whose deletion disconnects $G'$ into at least $K$ components. We
	can easily observe that $G'$ becomes disconnected only by removing
	vertices of $U$. Thus, the solution satisfies the condition that
	only vertices from $U$ are deleted. Indeed, if the condition
	is not satisfied, then $S$ should contain the original vertices of
	$G$ and/or vertices from the added paths $p_i$. Given such a
	solution an equivalent solution satisfying the condition that only
	vertices from $U$ are deleted can be constructed in polynomial
	time. In doing so, we swap each vertex $v \in S$ and a vertex $u
	\in U$, \emph{i.e.} we keep $v$ and we delete $u$ instead, and hence we
	get an induced graph with probably more components, since deleting vertices
	from $U$ can disconnect $G'$ and generates further components.
	Thus, the obtained solution is at least as good as $S$, and
	satisfies that only vertices from $U$ are deleted.
	
	Now, let $S \subseteq U$ be a solution of $I'$. In $I$, we select
	the $k$ edges of $A$ as follows: for each vertex $v \in S$, we
	select from $G$ the edge $uw \in E$ such that $uv,vw \in E'$. By
	deleting $A$ from $G$, no more than $k$ edges are deleted,
	$|A|\leq k$, and at least $K$ connected components are generated.
	Therefore, $A$ is a solution of the \emph{K-cut problem} on $G$.\\This
	complete the proof.
	
\end{proof}

\begin{remark}
It is clear that for the complete bipartite graph $K_{n_1,n_2}$ the
\emph{K-way vertex cut problem} is trivial, and the solution is
obtained by deleting the partition of smaller cardinality if $n_1, n_2
\leq k$. Otherwise, the solution is to delete any $k$ vertices
that results in only one component.
\end{remark}

\section{Split graphs}
\label{sec-sG}
Considering split graphs, we show that the \emph{K-way vertex cut problem} is equivalent to the \textit{Critical Node Problem} (\emph{CNP}) \cite{aru09}. \vspace{0.1cm}

\begin{theorem}
\label{splitg} The \emph{K-way vertex cut problem} and the CNP are equivalent on split
graphs.
\end{theorem}

\begin{proof}
Given a split graph $G=(V,E)$ and a set of vertices $S
\subseteq V$, we can easily notice that $G[V\setminus S]$ always
contains a non-trivial connected component and isolated vertices, if
any (see \textit{Figure \ref{proof_split}}). Note that $G$ is a split graph if the set of
vertices can be partitioned into two subsets $V_1$ and $V_2$, $V=V_1
\cup V_2$, where $V_1$ is an independent set and $V_2$ is a clique.
\\We recall that the recognition version of both the \emph{CNP} and \emph{K-way vertex cut problem} seeks for finding a
set of vertices of at most $k$, the deletion of which, respectively, minimizes
pairwise connectivity (for the \textit{CNP}), or maximizes the number of components (for the \textit{K-way vertex cut problem}) in
the remaining graph. According to the value of $k$, two cases can
be considered:

\begin{figure}[htbp]
	\centerline{\includegraphics[scale=0.18]{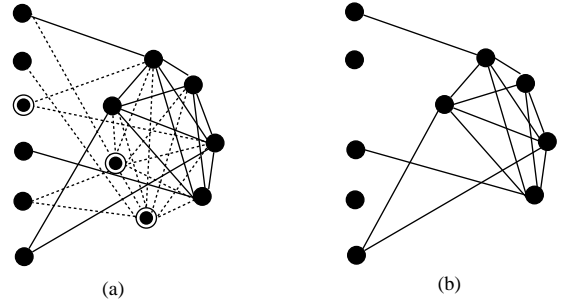}}
	\caption{Deleting any subset of vertices (eg. vertices with circles) from a split graph (see (a)) results in a non-trivial connected component and isolated vertices (see (b)).} \label{proof_split}
\end{figure}

\emph{Case 1:} \emph{$k\geq |N(V_1)|$}. This is a trivial case,
where the optimal solution, for both variants, is to delete the
vertices of $N(V_1)$ and any $k-|N(V_1)|$ vertices from $V_2$. We then
obtain a residual graph that has $|V_1|$ isolated vertices and a
connected component of size $|V_2|-(k - |N(V_1)|)$.

\emph{Case 2:} \emph{$k < |N(V_1)|$}. In this case, we consider an
optimal solution for the \emph{CNP} and try to prove that it is
also an optimal solution for the \emph{K-way vertex cut problem}, and vice versa. Given
an optimal solution $s^*$ for the \emph{CNP} on a split graph $G$,
this solution aims to find a set of vertices $S \subseteq V$ so that
the non-trivial connected component of $G[V \setminus S]$ is as
small as possible and the surviving isolated vertices of the
independent set be as large as possible. Therefore, we note that
for $s^*$ only vertices in $V_2$ are removed from $G$ (\emph{i.e.,}
$S \subseteq V_2$), and given any optimal solution for the
\emph{CNP}, an equivalent solution satisfying this condition
($S \subseteq V_2$) can be constructed in polynomial time (for
proof see \cite{add13}). On the other hand, to solve the \emph{K-way vertex cut problem}
we aim to obtain a maximal number of components in the residual
graph. In doing so, we seek for maximizing the number of isolated
vertices from $V_1$ once the critical vertices have been deleted. For
this purpose, only vertices in $V_2$ are removed from $G$, which is
exactly the solution $s^*$. Hence, the solution $s^*$ is also the
optimal solution of the \emph{K-way vertex cut problem}.
\\Therefore, an optimal solution of one of the two problems is an
optimal solution of the other, and so the \emph{CNP} and the
the \emph{K-way vertex cut problem} are equivalent.
\end{proof}

According to \emph{Theorem \ref{splitg}} and since the \emph{CNP}
is NP-complete on split graphs \cite{add13}, we have the following
corollary:\vspace{0,2cm}

\begin{corollary}
	The \emph{K-way vertex cut problem} remains NP-complete on split graphs.
\end{corollary}

\vspace{0,3cm}
This is also what has been proven by Berger \emph{et al.} \cite{ber14}
through a reduction from the k-clique problem.

\section{Other results}
\label{sec-btw}

We mentioned above that the \textit{K-way vertex cut problem} is polynomially solvable on graphs of bounded treewidth \cite{ber14}. The considered graphs are unweighted. In this section, we deduce that it remains polynomially solvable on the case of weighted graphs with bounded treewidth. Weighted graphs means that a weight $w_i \geq 0$ is associated with each node $ v_i \in V$. In this case, we ask for a subset of nodes of a total weight (rather than a cardinality) no more than $k$, whose removal maximizes the number of connected components in the induced graph.

In \cite{add13}, authors studied the \emph{MaxNumSC} problem, for \emph{Maximizing the Number of Small
	Components}, that can be formulated as
follows. Let $f^c(S)$ be the function that returns the number
of connected components in $G[V\setminus S]$ with a cardinality of
at most $c$. 
:\vspace{0.3cm}

\textbf{Input:} A graph $G = (V, E)$, and two integers $c$ and
$k$.

\textbf{Output:} $\underset{S \subseteq V}{arg max}$
$f^c(S)$, where $|S|\leq k$.\vspace{0.3cm}

Given a graph $G=(V,E)$ and two positive integers $k$ and $c$, the \emph{MaxNumSC} problem consists in maximizing the number of connected components of cardinality at most $c$, by deleting $k$ vertices from $G$.  The
authors showed that the problem is polynomially solvable on weighted
graphs with bounded treewidth. 

It is obvious that the \emph{K-way vertex cut problem} is a special case of the
\emph{MinMaxSC} problem where $c=|V|$, and as the
\emph{MinMaxSC} problem is polynomially solvable on weighted
graphs (where $w_i \geq 0, \forall v_i \in V$) we have the following corollary.\vspace{0.2cm}

\begin{corollary}
	The \emph{K-way vertex cut problem} is polynomially solvable on weighted graphs with bounded treewidth.
\end{corollary}

\section{Conclusion and future works}
\label{sec-conc}

In this paper, we studied the complexity of the \emph{K-way vertex cut problem} on some particular classes of graphs, namely 
bipartite and split graphs. This problem asks for finding the subset of
vertices in a graph, the deletion of which results in the maximum
number of connected components in the induced subgraph. We proved its
NP-completeness on bipartite graphs. While on split graphs, we provided its equivalence to the well-known problem, namely the \textit{Critical Node Problem} (\textit{CNP}). This allows any solving method for the \textit{CNP} to be used for solving The \textit{K-way vertex cut problem} and vice versa. 

The problem still needs more investigation on both complexity study and solving methods. For future works, we can consider it on subclasses of (or related classes to) bipartite and split graphs, which can help providing bounds for the problem hardness. In fact, this is what we are already started to do by considering  bipartite-permutation graphs (which is a subclass of the bipartite graph class). We found that the problem can be solved polynomially, on this class of graphs, using dynamic programming approach. Also, the problem can be investigated on other important classes of graphs, such as chordal graphs, disk graphs, etc. which will allow us to find different applications of this parameter in real-world networks.


\end{document}